\documentclass[11pt]{article}
\usepackage[margin=1in]{geometry}
\usepackage{graphicx} 
\usepackage[utf8]{inputenc}
\usepackage[font=small,labelfont=bf]{caption}
\usepackage{graphicx}
\usepackage{float}
\usepackage{subcaption}
\usepackage[numbers,sort&compress,elide]{natbib}
\usepackage{bm}
\usepackage{ulem}
\usepackage{hyperref}
\usepackage{authblk}  
\usepackage{upgreek}

\hypersetup{colorlinks=true,citecolor=blue, linkcolor=blue,urlcolor=blue}
\graphicspath{{figures/}}   

\begin{document}
\title{Quantum control of Nitrogen-Vacancy spin in Diamonds: Towards matter-wave interferometry with massive objects}
\author[1]{N. Levi}
\author[1]{O. Feldman}
\author[1]{Y. Rosenzweig}
\author[1]{D. Groswasser}
\author[2]{A. Elgarat}
\author[2]{M. Gal-Katizri}
\author[1]{R. Folman}

\affil[1]{Ben-Gurion University of the Negev, Department of Physics and Ilse Katz Institute for Nano-scale Science and Technology, Be'er Sheva 84105, Israel}
\affil[2]{Ben-Gurion University of the Negev, School of Electrical Engineering, Be'er Sheva 84105, Israel}
\maketitle

\begin{abstract} 
Quantum mechanics (QM) and General relativity (GR), also known as the theory of gravity, are the two pillars of modern physics. A matter-wave interferometer with a massive particle can test numerous fundamental ideas, including the spatial superposition principle---a foundational concept in QM---in previously unexplored regimes. It also opens the possibility of probing the interface between QM and GR, such as testing the quantization of gravity. Consequently, there exists an intensive effort to realize such an interferometer. While several approaches are being explored, we focus on utilizing nanodiamonds with embedded spins as test particles which, in combination with Stern–Gerlach forces, enable the realization of a closed-loop matter-wave interferometer in space-time. There is a growing community of groups pursuing this path \cite{White_paper_Morley2025}. We are posting this technical note (as part of a series of seven such notes), to highlight our plans and solutions concerning various challenges in this ambitious endeavor, hoping this will support this growing community. Here we present our work on quantum control of a nitrogen-vacancy spin system in bulk diamonds and in levitated diamonds as a step towards Stern-Gerlach interferometry with levitated nanodiamonds. Our simulations show that the current state of the art for spin coherence time in nanodiamonds of a few tens of microseconds, is good enough to enable an SGI spatial splitting on the order of nanometers for an ND composed of $10^7$ atoms. We would be happy to make available more details upon request.
\end{abstract}

\section{Introduction}

Quantum Mechanics (QM) is a pillar of modern physics. It is thus imperative to test it in ever-growing regions of the relevant parameter space. A second pillar is General relativity (GR), and as a unification of the two seems to be eluding continuous theoretical efforts, it is just as imperative to experimentally test the interface of these two pillars by conducting experiments in which foundational concepts of the two theories must work in concert.\\
 
The most advanced demonstrations of massive spatial superpositions have been achieved by Markus Arndt's group, reaching systems composed of approximately 2,000 atoms\,\cite{fein_quantum_2019}. This will surely grow by one or two orders of magnitude in the near future. An important question is whether one can find a new technique that would push the state of the art much further in mass and spatial extent of the superposition. Several paths are being pursued \cite{Romero-Isart_2017,Pino_2018,Weiss2021,kialka_roadmap_2022,Neumeier2024} and we choose to utilize Stern-Gerlach forces.\\

The Stern-Gerlach interferometer (SGI) has, in the last decade, proven to be an agile tool for atom interferometry \cite{amit_t3_2019,Keil2021,dobkowski_observation_2025}. Consequently, we, as well as others, aim to utilize it for interferometry with massive particles, specifically, nanodiamonds (NDs) with a single spin embedded in their center \cite{scala_matter-wave_2013,wan_free_2016,margalit2021_OUR_intro}.
Levitating, trapping and cooling of massive particles, most probably a prerequisite for interferometry with such particles, has been making significant progress in recent years. Specifically, the field of levitodynamics is a fast growing field \cite{GonzalezBallestero2021Levitodynamics}. Commonly used particles are silica spheres. As the state of the art spans a wide spectrum of techniques, achievements and applications, instead of referencing numerous works, we take, for the benefit of the reader, the unconventional step of simply mapping some of the principal investigators; these include Markus Aspelmeyer, Lukas Novotny, Peter Barker, Kiyotaka Aikawa, Romain Quidant, Francesco Marin, Hendrik Ulbricht and David Moore. Relevant to this work, a rather new sub-field which is now being developed deals with ND particles, where the significant difference is that a spin with long coherence times may be embedded in the ND. Such a spin, originating from a nitrogen-vacancy (NV) center, could enable the coherent splitting and recombination of the ND by utilizing Stern-Gerlach forces \cite{scala_matter-wave_2013,wan_free_2016,margalit2021_OUR_intro}. This endeavor includes principal investigators such as Tongcang Li, Gavin Morley, Gabriel Hetet, Tracy Northup, Brian D’Urso, Andrew Geraci, Jason Twamley and Gurudev Dutt.\\
 
We aim to start with an ND of $10^7$ atoms and extremely short interferometer durations. Closing a loop in space-time in a very short time is enabled by the strong magnetic gradients induced by the current-carrying wires of the atom chip \cite{Keil2016}. Such an interferometer will already enable to test the existing understanding concerning environmental decoherence (e.g., from blackbody radiation), and internal decoherence \cite{henkel_universal_2024}, never tested on such a large object in a spatial superposition. As we slowly evolve to higher masses and longer durations (larger splitting), the ND SGI will enable the community to probe not only the superposition principle in completely new regimes, but in addition, it will enable to test specific aspects of exotic ideas such as the Continuous spontaneous localization hypothesis \cite{adler_continuous_2021, gasbarri_testing_2021}. As the masses are increased, the ND SGI will be able to test hypotheses related to gravity, such as modifications to gravity at short ranges (also known as the fifth force), as one of the SGI paths may be brought in a controlled manner extremely close to a massive object \cite{geraci_short-range_2010, geraci_sensing_2015, bobowski_novel_2024, panda_gravitational_2024}. Once SGI technology allows for even larger masses ($10^{11}$ atoms), we could test the Diósi–Penrose collapse hypothesis \cite{Bassi_2013,penrose_on_2014, fuentes_quantum_2018, howl_exploring_2019, tomaz_gravitationally-induced_2024} and gravity self-interaction \cite{hatifi_revealing_2020, grossardt_dephasing_2021, aguiar_probing_2024} (e.g., the Schrödinger-Newton equation). Here starts the regime of active masses, whereby not only the gravitation of Earth needs to be taken into account. Furthermore, it is claimed that placing two such SGIs in parallel will allow probing the quantum nature of gravity \cite{bose_spin_2017, marletto_gravitationally-induced_2017}. This will be enabled by ND SGI, as with $10^{11}$ atoms the gravitational interaction could be the strongest \cite{vdKamp_quantum_2020, schut_relaxation_2023, schut_micrometer-size_2024}.\\
 
Let us emphasize that, although high accelerations may be obtained with multiple spins, we consider only a ND with a single spin as numerous spins will result in multiple trajectories and will smear the interferometer signal. We also note that working with an ND with less than $10^7$ atoms is probably not feasible because of two reasons. The first is that NVs that are closer to the surface than 20\,nm lose coherence, and the second is that at sizes smaller than 50\,nm, the relative fabrication errors become large, and a high-precision ND source becomes beyond reach. \\

In the past we have done extensive work with NV centers\,\cite{waxman_diamond_2014,schlussel_wide-field_2018,rosenzweig_probing_2018}. Here we describe our basic techniques for coherent control of the NV-center spin in bulk and in levitated NDs. We also describe our roadmap for the near future. This technical note is part of a series of seven technical notes put on the archive towards the end of August 2025, including a wide range of required building blocks for the ND SGI \cite{Liran_ND_neutralization, Benjaminov_UHV_ND_loading, Givon_ND_fabrication, Skakunenko_Needle_trap, Feldman_Paul_trap_ND, Muretova_ND_theory}.

\section{Background}

The NV color center in diamonds is the heart of the ND SGI experiment. This negatively charged defect in the diamond lattice exhibits the longest coherence time observed in a solid-state device at room temperature. In bulk diamonds, at cryogenic temperatures the T$_2$ time of the NV$^{-}$ can exceed one second, by using dynamical decoupling sequences \cite{bar-gill_solid-state_2013,abobeih_one-second_2018}. However, at room temperature this value drops to only a few milliseconds in $^{12}$C purified diamonds. In NDs the T$_2$ values are shorter, for example in milled NDs, the NVs T$_2$ time can exceed 400\,$\upmu$s at room temperature by applying dynamical decoupling \cite{wood_long_2022}. In milled NDs from polycrystalline $^{12}$C, a T$_2$ time of up to 786\,$\upmu$s was observed by using a spin-locking pulse sequence \cite{march_long_2023}.\\

Our route towards massive-object interferometry is based on a single NV embedded in a ND. The long coherence time of the NV spin system allows preparing it in a spin superposition, realizing thereafter spatial splitting and recombination using a SG scheme \cite{amit_t3_2019,dobkowski_observation_2025,Keil2021,margalit2021_OUR_intro}. Nonetheless, for a finite magnetic gradient, the magnitude of the splitting depends on the mass of the ND. Here, several factors come into play. If the ND is heavy the splitting could be too small, and on the other hand, in small NDs ($<$\,100 nm), the coherence time of the NV is reduced primarily due to interactions of the shallow NV with unpaired surface spins, dangling bonds, and charge fluctuations at the surface \cite{zhang_depth-dependent_2017, sangtawesin_origins_2019}. This will limit the interferometry time (requiring sufficient coherence times for both the spatial dimension and spin) and restrict the splitting. Even when using the much stronger diamagnetic forces (also originating from magnetic gradients) which have been shown to enable much larger splitting when applied differentially to the two wavepacket trajectories once an initial spin-based splitting has been achieved, the spin coherence time is the limiting factor when large spatial splitting is sought after. Other decoherence mechanisms, such as environmental spatial decoherence, e.g., from blackbody radiation, have been calculated by us to enable large spatial splitting, even at room temperature.\\

The first ND SGI experiments are planned to include NDs with about $10^7$ atoms (a few tens of nanometer diameter) and an SGI duration of a few tens of microseconds. With the available current densities on the atom chip giving rise to magnetic gradients as high as $10^5$\,T/m, this will already enable a splitting on the order of nanometers. To increase the coherence in the SGI, it is possible to utilize the Hahn-Echo technique. The group of Dirk Englund reports a Hahn-Echo T$_2$ time of 79\,$\upmu$s in CVD etched ND pillars with a diameter of 50\,$\pm$\,15\,nm and height of 150\,$\pm$\,75\,nm\,\cite{trusheim_scalable_2014}, while the group of Gavin Morley reports similar coherence times\,\cite{march_long_2023}. Such a coherence time is already sufficent for our first ND SGI runs. It is expected that with a high $^{12}$C purity sample the T$_2$ time will be even longer.\\

In levitated NDs, quantum control and measurements of the spin state of the NV are more complex, as due to rotations and translations of the ND, the orientation of the NV's axis is not fixed relative to the lab frame. As a result, the Zeeman level splitting due to the interaction with the external magnetic field constantly changes and the strength of the interactions with an external microwave field that drives the hyperfine transitions of the NV also changes continuously due to varying detuning.\\ 

Elongated NDs can have a naturally preferred orientation in a Paul trap \cite{perdriat_rotational_2024}. This “alignment stabilization” arises from the torque exerted by the trap’s electric field on the charged ND. More specifically, for charged non-spherical particles (like diamonds or graphene flakes), torque arises via coupling to the RF quadrupole field. The torque strength typically grows with the particle’s quadrupolar moment, which increases with volume and shape anisotropy. Hence, in the Paul trap the ND will naturally align with its long axis in the direction of the axis of the trap with the lower electric gradient.
The group of Gabriel H\'etet used microdiamonds with a diameter d\,$>$\,8\,$\upmu$m to lock the orientation of the diamond in the trap\,\cite{voisin_nuclear_2024}. This allowed them to further rotate the external magnetic field such that one of the possible four NV orientations in the crystal will be aligned with it. Under such conditions they were able to achieve nuclear spin polarization and improvement of the coherence times by a factor that is close to the ratio of nuclear and electronic spin gyromagnetic factors.\\

The group of Tongcang Li demonstrated alignment of NDs using a chip Paul trap and additional electrodes which could be operated to actively rotate the ND at up to 20\,MHz and lock its orientation in the lab frame by gyroscopic stabilization\,\cite{li_quantum_2024}. Such a gyroscopic locking will considerably relax the requirements for rotation cooling\,\cite{wachter_gyroscopically_2025,rizaldy_rotational_2024} (See also our work in \cite{japha_quantum_2023,Muretova_ND_theory}). This was achieved under high-vacuum conditions which opens the possibility for implementing similar gyroscopic stabilization in the ND SGI.\\

\section{Experimental Setup}
\begin{figure}[ht!]
\centering
\includegraphics[width=\textwidth,trim=0mm 0mm 0mm 0mm,clip]{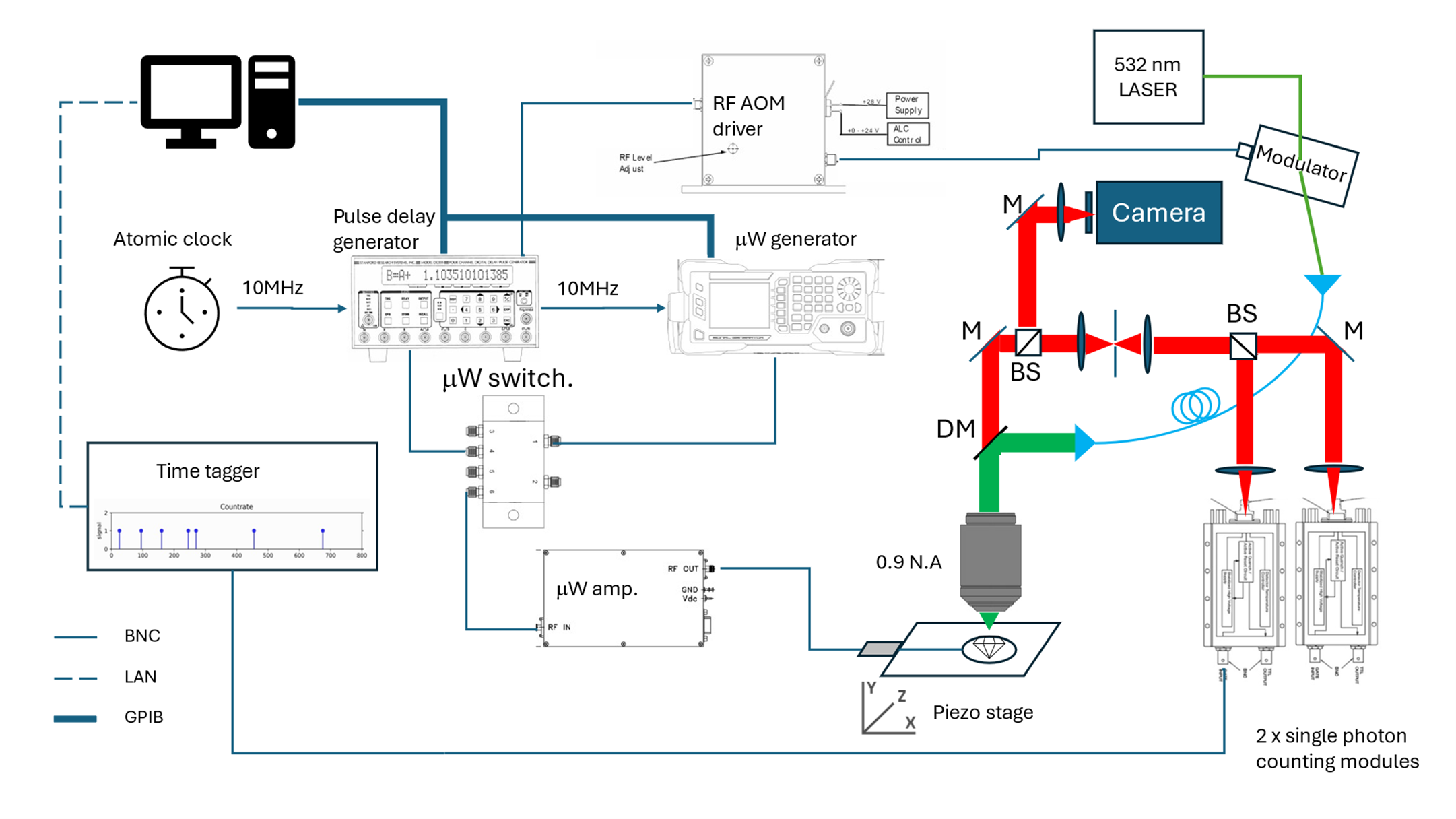}\vspace{-0.2cm}
\caption[Schematic diagram]{A schematic diagram of the experimental setup for measuring the NV spectroscopy by Optically Detected Magnetic Resonance (ODMR) and for quantum control of the NV spin system. The design is based on a confocal fluorescence microscope (CFM) architecture. The Red fluorescence from the NV is separated from green reflections using a dichroic mirror (DM) and is directed by mirrors (M) and 50:50 beam splitting cubes (BS) towards a CCD camera for imaging and two single photon counting (SPCM) units for detection and g$^{(2)}$ measurements. We use a telescope and a pinhole of 15\,$\upmu$m to further isolate and localize the spatial origin of NV fluorescence.}
\label{fig:System.png}
\end{figure}

Our confocal fluorescence microscopy (CFM) setup for the characterization of different samples of NVs in diamonds is depicted in Fig.\,\ref{fig:System.png}.\\

The experimental apparatus consists of a confocal fluorescence microscope (CFM) design. The experimental control is programmed in Python and all the “real time” operations are triggered by TTLs from a pulse-delay-generator (PDG) (SRS DG645). A Rb atomic clock (SRS SR625) is used to provide a stable 10\,MHz time base to the PDG and to the microwave generator (Rigol DSG836A) which is used to drive the hyperfine transitions of the NV. The PC communicates with the microwave generator and the PDG by GPIB commands to execute asynchronous commands, such as changing the microwave frequency during optically detected magnetic resonance (ODMR) spectrum or changing the duration of the microwave pulses during a Rabi sequence. The microwave chain includes a microwave switch (Mini Circuits ZASWA-2-50DR+) to pulse the microwave power, which is regulated by the PDG TTL output. The microwave power is amplified (Mini Circuits ZHL-16W-43-S+) before it excites a microwave antenna on a chip (see Appendix A). A diamond is placed in the middle of the antenna and is brought to the focus of the 0.9 N.A. microscope objective (OLYMPUS LMPlan Apo 150X) using an x,y,z stage that is controlled by piezo actuators (ThorLabs DRV517). We use a 532\,nm laser to excite our NVs (Coherent® Sapphire™ LPX 500). In measurements of the ODMR signal, the laser is continuously on, and the fluorescence count rate is measured as a function of the microwave frequency (Fig.\,\ref{fig:ODMR.png}). To perform optical pumping and for the readout of the NV spin state, the laser power is pulsed by an Acusto-Optic modulator (AOM) (Crystal Technologies 3080-110). This AOM has a a rise/fall time smaller than 25 \,ns given that the beam is focused properly. The AOM is powered by an RF driver (Crystal Technology 1080AF-AENO-2.0) which is regulated using the PDG. In practice, we are able to achieve rise/fall times of 17\,ns, which is important for the data taking, as the metastable singlet state of the NV have a lifetime on the order of only 300\,ns (see Fig.\,\ref{fig:Normalised_PL.png}). The first order diffraction from the AOM is coupled into a polarization maintaining optical fiber. Overall we achieve an extinction ratio larger then 1,000. The output power (10-20\,mW) from the fiber is collimated and reflected to the microscope objective via a dichroic mirror (Semrock) which also separates the red fluorescence from the NV on the path to the detection optics which includes a two SPCM units (SPCM-AQRH14) and a COMS camera (Ueye SE). The signals from the SPCM are transferred to a time tagger (Time Tagger 20, Swabian Instruments), which is connected to the PC via LAN for the statistical data acquisition.

\section{Results in bulk}
\begin{figure}[ht!]
\centering
\includegraphics[width=1\textwidth,trim=0mm 0mm 0mm 0mm,clip]{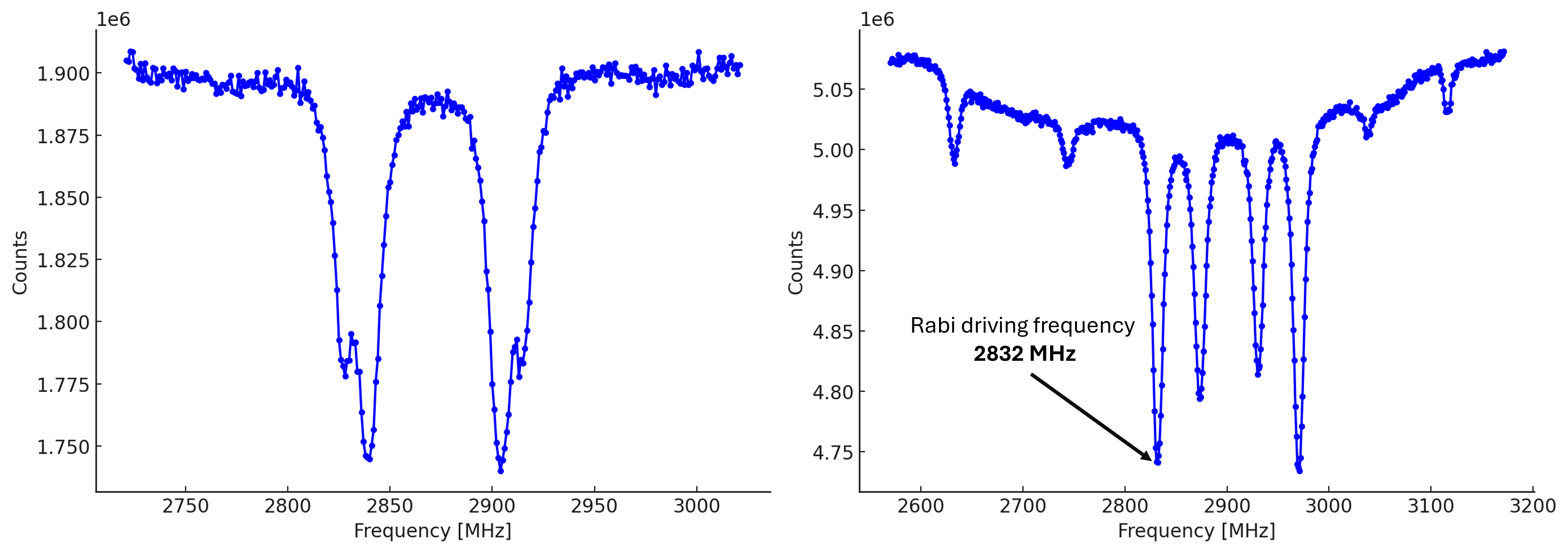}\vspace{-0.2cm}
\caption[ODMR]{ODMR performed on a bulk diamond (EL SC Plate 2.0\,mm x 2.0\,mm, 0.50\,mm thick $\{$100$\}$) with different magnetic fields. (Left) A magnet was placed under the sample so that the four different crystallographic orientations of the NV are almost symmetric relative to the direction of the magnetic field and thus the Zeeman splitting are almost perfectly degenerate. (Right) The magnetic field is oriented asymmetrically with respect to the four NV orientations, leading to four distinct Zeeman splittings. The depth of the spectroscopy dips is due to the angle with the magnetic field of the microwave radiation and the proximity to the focal point of the objective, as there are very few NV centers in the focal point. The Rabi driving frequency of 2832 MHz, shown on the graph, is used for the Rabi measurement in Fig.\,\ref{fig:Rabi.png}.}
\label{fig:ODMR.png}
\end{figure}

\begin{figure}[H]
\centering
\includegraphics[width=0.8\textwidth,trim=0mm 0mm 0mm 0mm,clip]{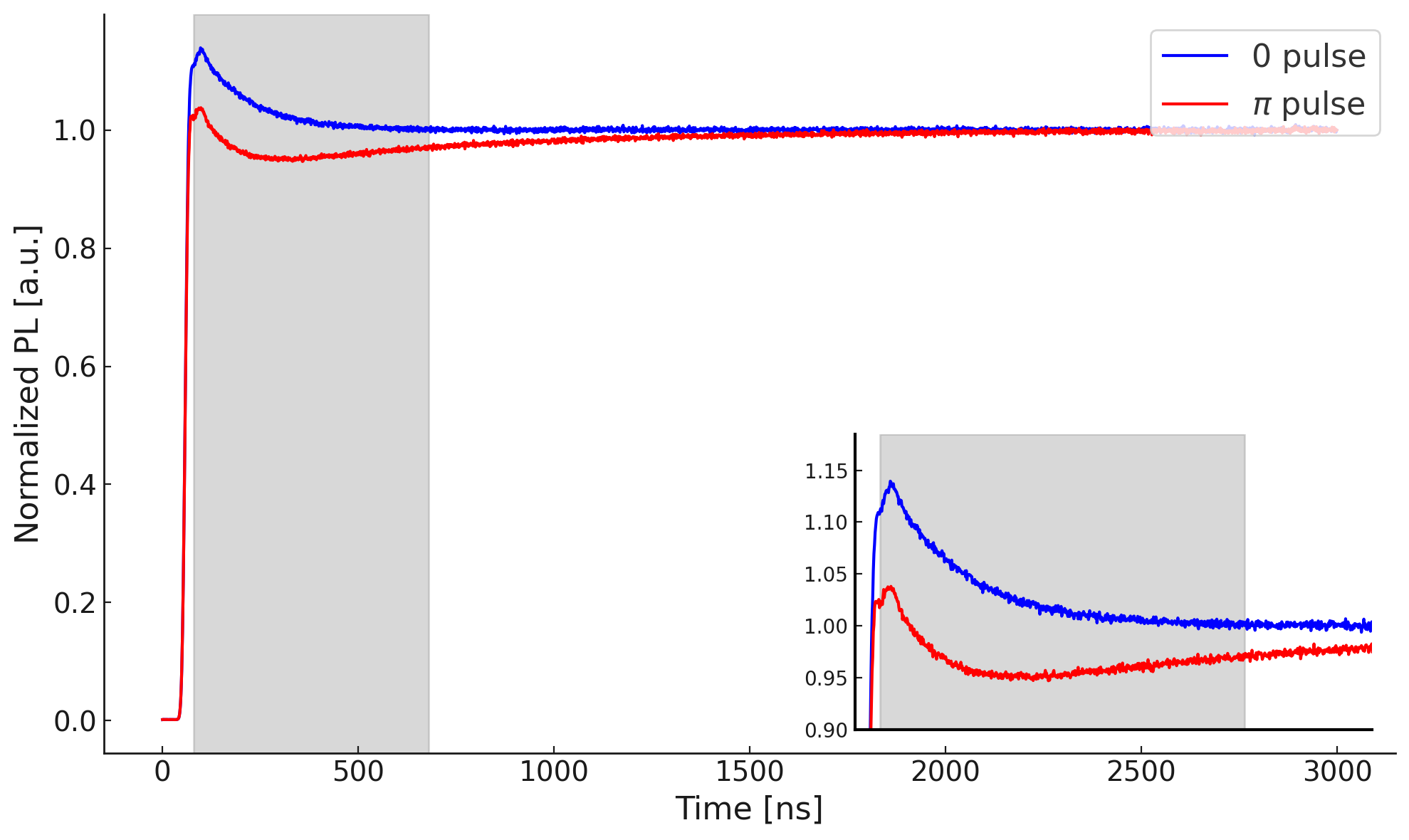}\vspace{-0.2cm}
\caption[Normalized fluorescence]{Normalized fluorescence response for the spin states m$_s$\,=\,0,\,$\pm$\,1. A long 532\,nm laser pulse pumps the NV to the bright m$_s$\,=\,0 state (blue trace). If optical pumping is followed by a resonant microwave $\pi$ pulse (red trace), a transient fluorescence response reveals the dynamics of the NV system (see inset for zoom in) and is used for readout of the NV state. When the dynamics reaches a steady state, the optical pumping to the bright m$_s$\,=\,0 state is complete.
Typical relaxation times to the m$_s$\,=\,0 state are around 300\,ns. Here the relaxation time is much longer, and the reason is yet unclear
\cite{hopper_spin_2018}. In the ND SGI, the final spin readout, which is the observable of the interferometer, will be dark, namely, without spectroscopy. This is done by a SG magnetic gradient pulse which spatially separates the different spin states until a CCD can resolve them. This is the method we have used in our atomic SGI, and it has a much better fidelity than spectroscopic readouts, and in addition, does not internally heat the ND.}
\label{fig:Normalised_PL.png}
\end{figure}

\begin{figure}[H]
\centering
\includegraphics[width=0.7\textwidth,trim=0mm 0mm 0mm 0mm,clip]{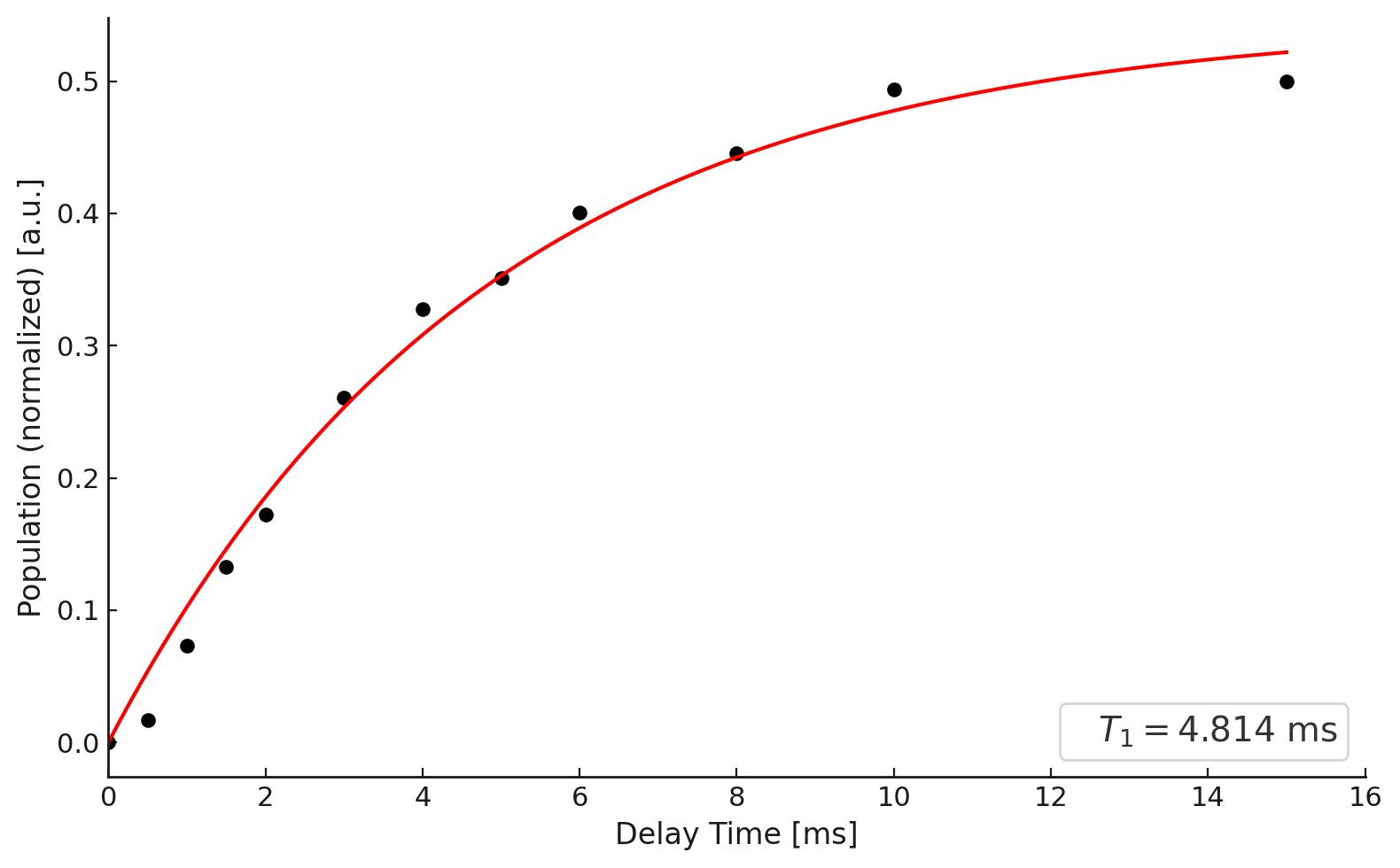}\vspace{-0.2cm}
\caption[T1]{A T1 measurement. The sequence includes an optical laser pulse which is used for readout/optical pumping which is followed by free evolution in the dark for a duration $\tau$. This sequence is repeated many times to obtain a high signal/noise ratio.}
\label{fig:T1.jpg}
\end{figure}

In Fig.\,\ref{fig:ODMR.png} we demonstrate ODMR performed on a $\{$100$\}$ diamond from element six (with a NV concentration of less than 5\,ppb). Once the resonances are detected it is possible to control the spin state of the system. For this, the NV system is initialized in the m$_s$\,=\,0 state by optical pumping using a laser pulse. Following initialization, the system can evolve undisturbed in the dark or be subjected to a microwave pulse. In Fig.\,\ref{fig:Normalised_PL.png} we demonstrate the contrast that we observe following a $\pi$ pulse. In Fig.\,\ref{fig:T1.jpg} we show a T$_1$ measurement, where the NV system freely evolves in the dark for a time $\tau$ which is followed by another laser pulse. The beginning of each laser pulse is used for the readout of the spin state and the end of the pulse is used for optical pumping.\\

\begin{figure}[H]
\centering
\includegraphics[width=0.7\textwidth,trim=0mm 0mm 0mm 0mm,clip]{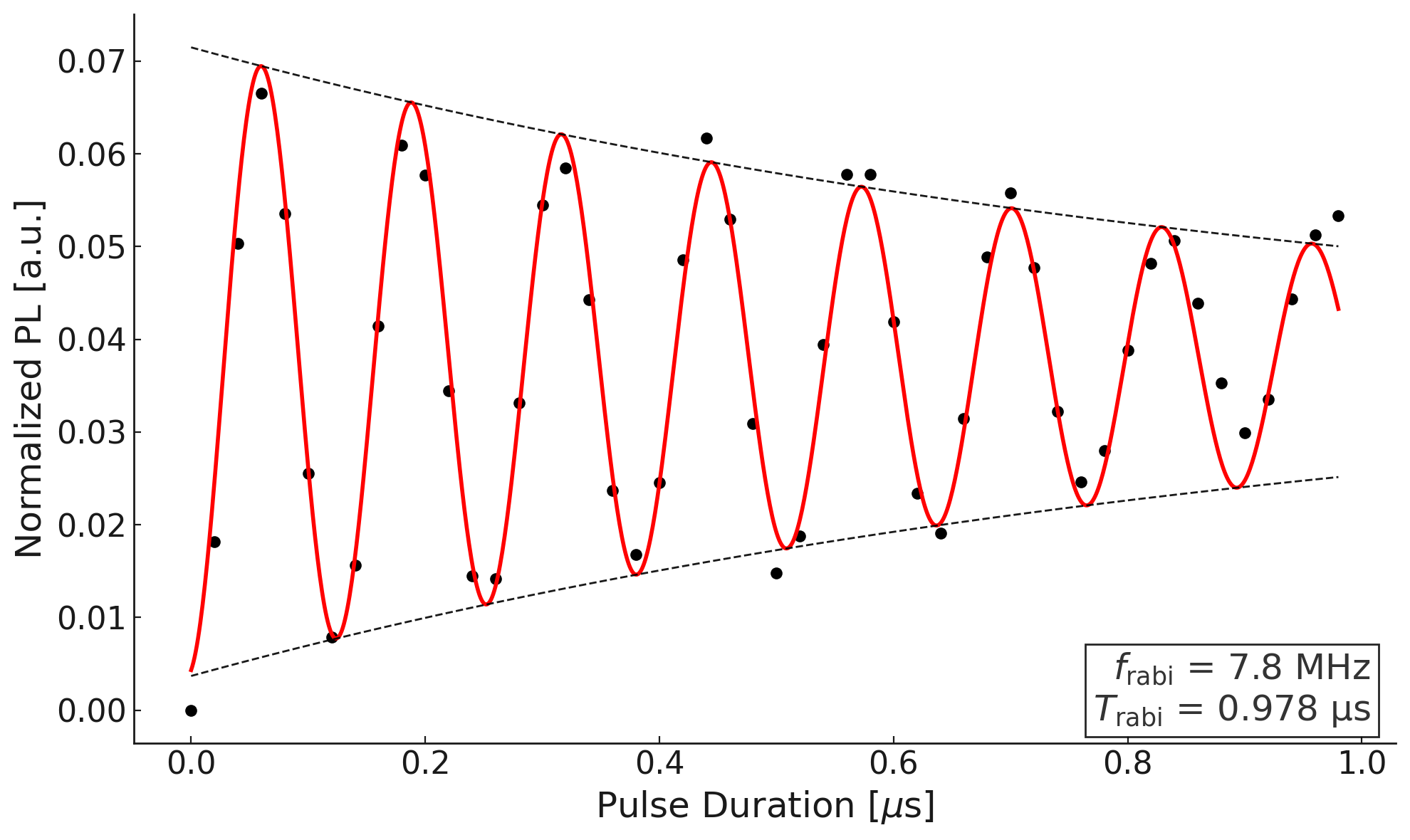}\vspace{-0.2cm}
\caption[Rabi]{A resonant Rabi oscillation on the same diamond as in Fig.\,\ref{fig:ODMR.png}.}
\label{fig:Rabi.png}
\end{figure}

\begin{figure}[H]
\centering
\includegraphics[width=0.7\textwidth,trim=0mm 0mm 0mm 0mm,clip]{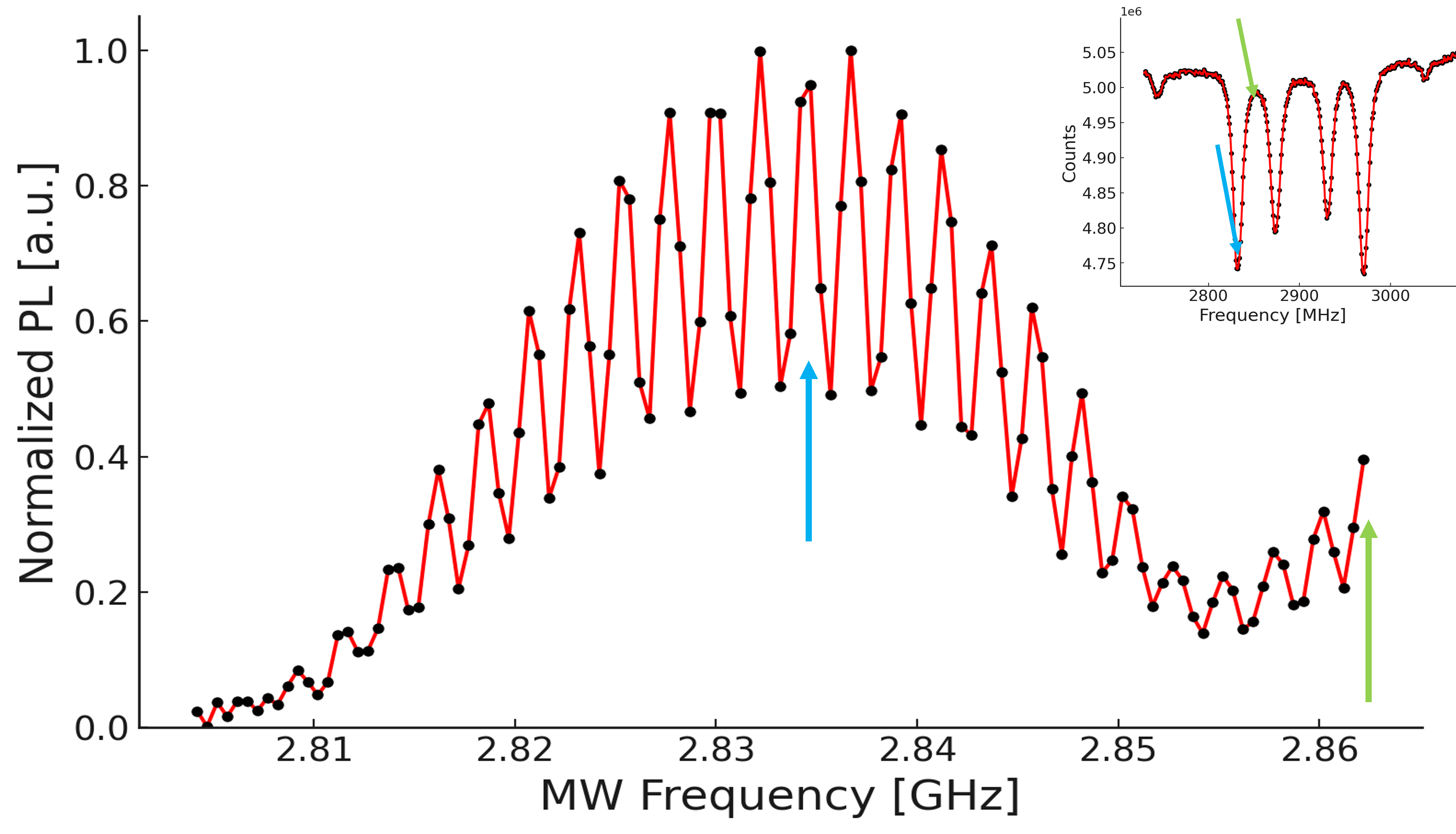}\vspace{-0.2cm}
\caption[RamseyFringes]{Ramsey fringes as a function of frequency around the resonant center frequency of 2832 MHz (blue arrows). 
The green arrows show 
that the rise of the fringe pattern on the far right is caused by the second resonant frequency at 2874 MHz. Inset: ODMR spectrum taken from Fig.\,\ref{fig:ODMR.png}.}
\label{fig:RamseyVsFreq.png}
\end{figure}

\begin{figure}[H]
\centering
\includegraphics[width=0.7\textwidth,trim=0mm 0mm 0mm 0mm,clip]{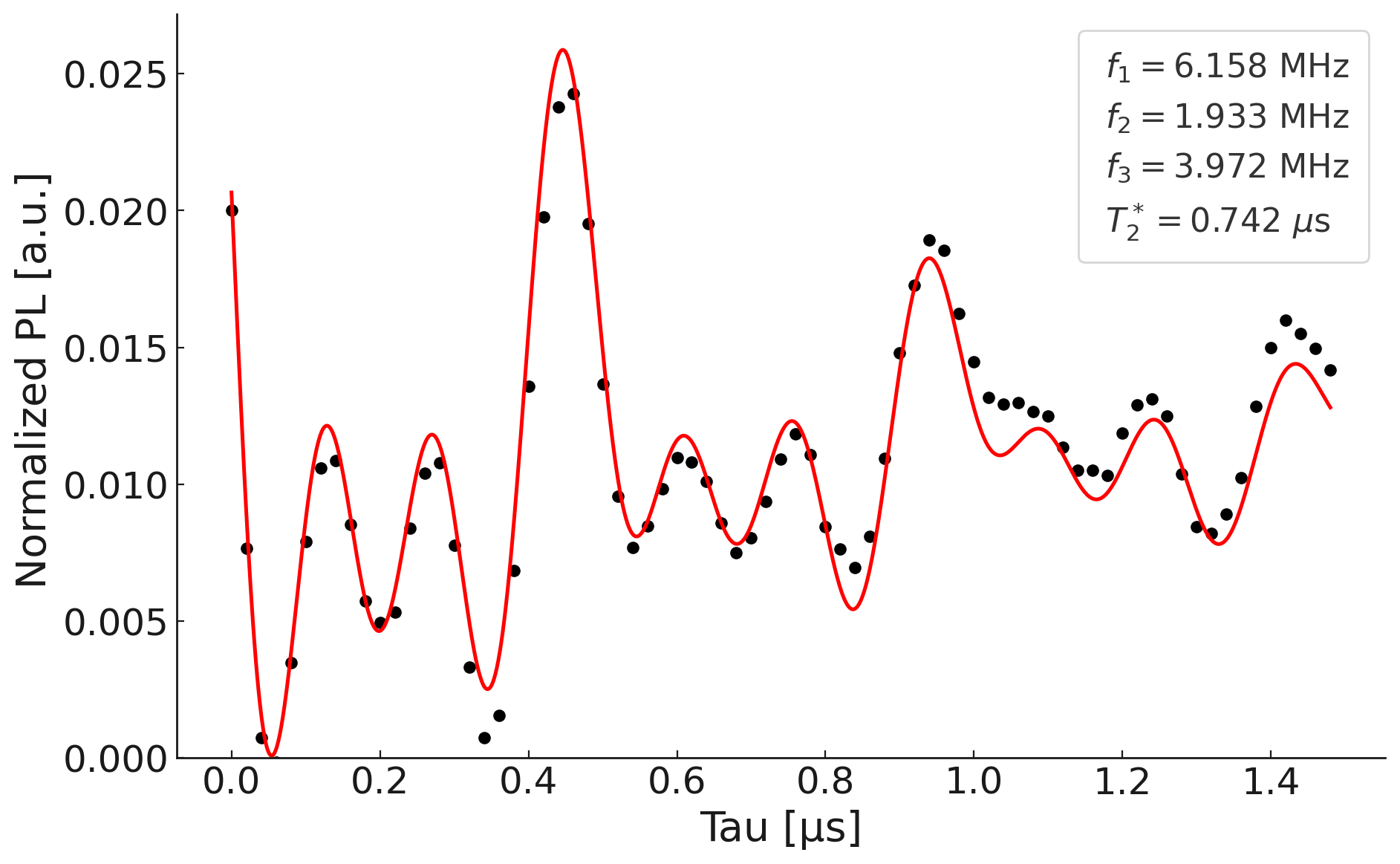}\vspace{-0.2cm}
\caption[Ramsey]{Ramsey oscillations as a function of time. The data is fitted to a sum of three cosine functions
\( f(t) = A + B e^{-t/T_2^*} \sum_{k=1}^{3} \cos\,\!\bigl(2\pi f_k t + \varphi_k\bigr) \),
as there are three closely spaced hyperfine transitions, each providing a different detuning (see Appendix~B). As shown in the inset, the fit returns three frequencies spaced by about \(2\,\mathrm{MHz}\), as expected by theory.}
\label{fig:RamseyVsTime.png}
\end{figure}

Finally, in Figs.\,\ref{fig:RamseyVsFreq.png} and \ref{fig:RamseyVsTime.png} we present Ramsey oscillations as a function of frequency and as a function of time.

\section{ODMR on levitated diamonds in a Paul trap}
\begin{figure}[ht!]
\centering
\includegraphics[width=0.8\textwidth,trim=0mm 0mm 0mm 0mm,clip]{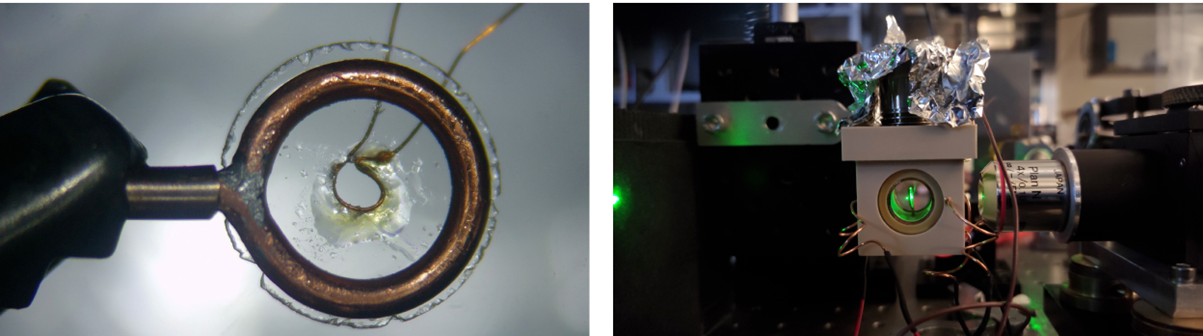}\vspace{-0.2cm}
\caption[LevSystem] {ODMR spectroscopy on levitated NDs: (Left) A testbed ring trap glued to a quartz plate with a central through-hole; a wire antenna on the reverse side provides the microwave field. (Right) A cluster of NDs levitated near the ring trap with a microwave antenna. The antenna’s ground potential shifts the RF null point of the ring trap. Consequently, the particle levitates outside the ring plane rather than at the geometric center. The eventual setup will obviously require a much more advanced antenna and we are developing such antennas, see Appendix A. In the first generation of the ND SGI, a chip with current-carrying wires giving rise to the SG magnetic gradient, and with an antenna layer, will be brought in close proximity to our existing macroscopic Paul traps \cite{Skakunenko_Needle_trap,Feldman_Paul_trap_ND}.}
\label{fig:Levitated.png}
\end{figure}

As a first test, we used a simple ring Paul trap inside a PEEK cube vacuum chamber to levitate the diamonds (see Fig.\,\ref{fig:Levitated.png}). The diamonds are loaded into the trap by the “stick” method where a drop of particle solution is pipetted onto a microscope slide and dried at 80$^{\circ}$\,C for about an hour. Using a scalpel, a thin powder layer is scraped, forming a powder on the slide. Next, we rub a PTFE (Teflon) rod with cotton or wool to build up a static charge on the rod. Holding the charged rod a few millimeters above the slide picked up some of the powder and charged the particles. Finally, we put the rod through the loading port of the vacuum chamber and gently shake it, so that charged particles fall into the ring Paul trap. We later used a system we developed which is based on the electro-spray method \cite{Feldman_Paul_trap_ND}, and in parallel are developing a dry method \cite{Benjaminov_UHV_ND_loading} to load particles to the trap in UHV.\\

The optical setup for obtaining the ODMR in levitated diamonds is basically the same as described above, but, in this setup we used a microscope objective (Olympus PLN 4X) with a long 18.5\,mm working distance and 0.1 N.A. to focus the green light on the diamond and collect the red fluorescence. \\

The trap is formed by a 2\,mm radius ring by bending a 0.8\,mm diameter wire. The ring is glued to a quartz plate with a through-hole for laser access. A simple wire antenna, shaped like an $\Omega$ and mounted on the opposite side of the plate, provides the MW field.\\

\begin{figure}[H]
\centering
\includegraphics[width=0.7\textwidth,trim=0mm 0mm 0mm 0mm,clip]{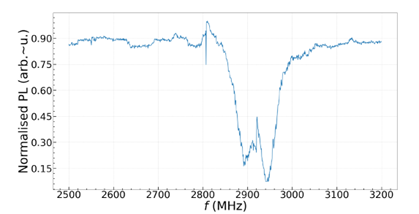}\vspace{-0.2cm}
\caption[LevODMR]{ODMR spectrum of a levitated ND in the Earth magnetic field. Two distinct resonance dips are visible, corresponding to the transitions to the m$_s$\,=\,+1 and to the m$_s$\,=\,-1 spin states.}
\label{fig:LevODMR.png}
\end{figure}

The ODMR spectrum presented in Fig.\,\ref{fig:LevODMR.png}, shows only two spectroscopy dips as no magnetic field was applied. These two dips are due to the lattice strain causing a zero-field splitting in the transition between m$_s$\,=\,0 and m$_s$\,=\,$\pm$\,1. Once we achieve alignment of the ND (through electrical alignment - see our preprint on a high-frequency Paul trap \cite{Skakunenko_Needle_trap}), we will be able to apply a magnetic field and observe the full eight-dip ODMR spectrum.

\section{Discussion}
In this technical note we described basic setups and methods used to measure and exploit the quantum properties of the NV system. This work is one building block of a bigger and ambitious plan to use NDs with a single NV in a SG interferometer (ND SGI). If successful, this project will push the mass for which a coherent spatial superposition is observed by orders of magnitude and will provide experimental data that is necessary for our understanding of some of the most fundamental questions such as the border between the quantum world and the classical world, and the interface between gravity and quantum mechanics.\\

The main goal is now to achieve a T$_2$ time of a few tens of $\upmu$s without any dynamic decoupling sequence (aside from Hahn-echo), which is good enough for a short-duration ND SGI. As this has already been achieved by others \cite{trusheim_scalable_2014,march_long_2023} (though, as far as we know, not in levitated NDs), we do not anticipate any fundamental problems.\\

In parallel to the effort described in this note, we are developing the integration of the experimental tools that were described here into the interferometer setup. Future work on the NV track will focus on conducting all the quantum operations on levitated NDs after orientation stabilization in our tighter Paul traps \cite{Skakunenko_Needle_trap}. While levitated NDs pose a challenge in terms of quantum operations in the context of rotations, they also present opportunities. Specifically, as noted, the single-shot readout becomes much easier, as it can be done by a SG pulse complemented with straightforward imaging and not through spectroscopy. This is the method we have used with our atomic SGI, and it promises much higher fidelity with much lower heating of the ND. 

\section{Outlook}

Several steps still need to be taken until the NV is fully controlled within a ND SGI experiment.
For example, while the presently achieved T$_2$ times are sufficient for short-duration SGI, for longer durations, longer coherence times will need to be achieved.
As is well known, dynamic decoupling has been successfully applied to prolong T$_2$ (see for example \cite{louzon_noise_2025} and references therein), but for SGI schemes one has to be careful that the dynamic-decoupling spin manipulation does not interfere with the SG forces.
Indeed, several works already serve as an indication that this will be possible \cite{pedernales_motional_2020, wood_spin_2022}.\\

\begin{figure}[H]
\centering
\includegraphics[width=1\textwidth,trim=0mm 0mm 0mm 0mm,clip]{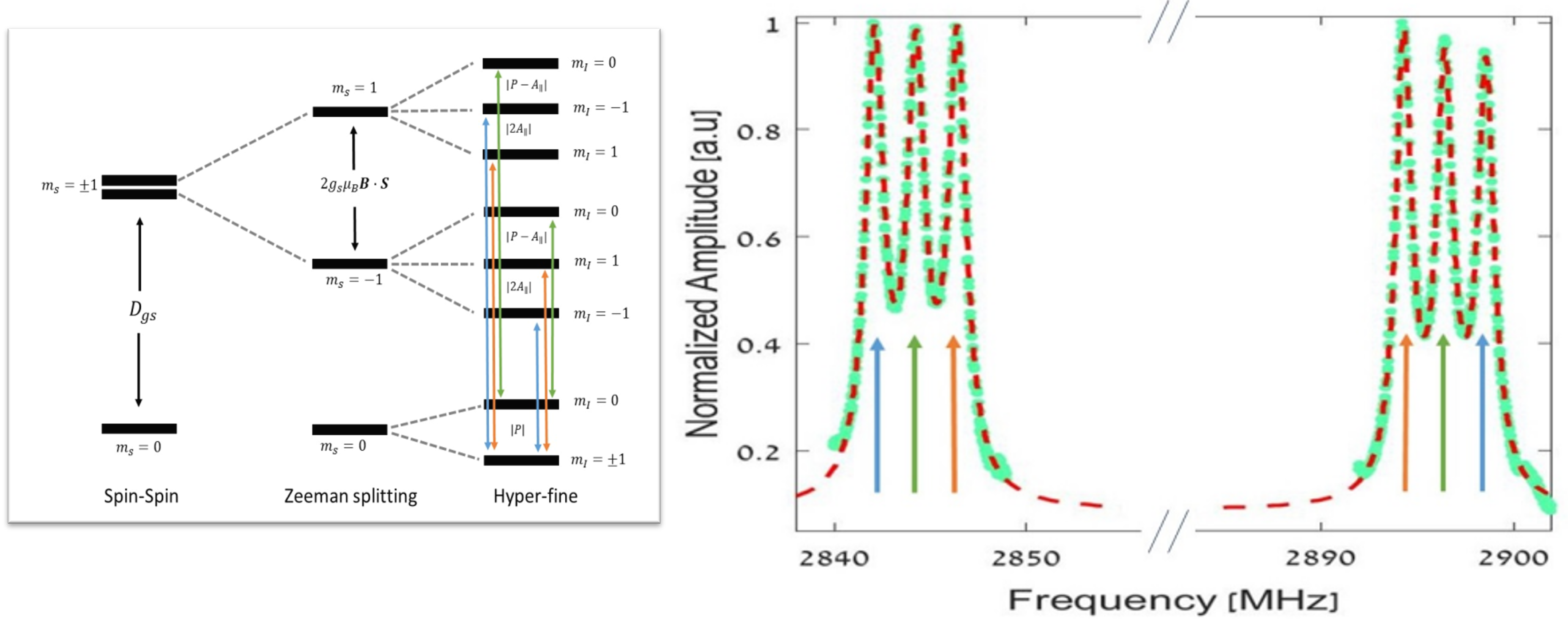}\vspace{-0.2cm}
\caption[Hyperfine] 
{(Left) Schematic of the $^{14}$N NV energy ground state levels arising from interactions splitting the ground state
in the presence of an axial magnetic field. The energy splittings are expressed in terms of the hyperfine
constants. Allowed transitions are indicated by double arrows. Electronic and nuclear spin projections
are denoted by m$_s$ and m$_I$, respectively. (Right) High resolution ODMR scan of the $\{$111$\}$ orientation with -12 dBm MW Power. Data in turquoise, fit of 6 Lorentzians in red. The two batches of three peaks represent the transitions from
the m$_s$ = 0 to the m$_s = \pm1$ states and each color of arrow represents a transition for a different m$_I$ state. the blue arrows represent transitions with m$_I = -1$ nuclear spin. Green and orange lines are the transitions with m$_I$ = 0 and m$_I$ = 1 respectively.}
\label{fig:Hyperfine_Splittings.png}
\end{figure}

Another common approach to improve coherence time is to map the electron spin states onto the nitrogen nucleus. While the latter nuclear states are not usable for SG forces, they are usable as a long-T$_2$ quantum memory which could then be mapped back to electron states. Hence for experiments testing gravitationally-induced entanglement between two NDs (as a way to probe the quantization of gravity), the electron spins may be used for short durations to enable SG splitting and recombination, while for the long duration required for creating entanglement via the gravitational interaction, the spin may be stored in the nuclear long-lifetime state. Recent experiments with the nitrogen nuclear states in levitated diamonds have already shown three orders of magnitude longer coherence time\,\cite{voisin_nuclear_2024}. In Fig.\,\ref{fig:Hyperfine_Splittings.png} we demonstrate our work on resolving the spectroscopy of the nuclear states.\\

Finally, while working with commercially available NDs is good enough  for the first generation of ND SGI, we are working
towards utilizing a high-precision source for NDs, in which the exact size and shape of the ND can be controlled, as well as its purity (See our preprint on fabrication \cite{Givon_ND_fabrication}). For example, we will start with an elongated box and dimensions on the order of 100\,nm ($10^7$ atoms) having a single NV in its center. 

\section{APPENDIX A.\,\,\, Microwave excitation of NV centers}

The negatively charged NV center in NDs is a promising system for large-mass spatial interferometry. Coherent manipulation of the NV’s spin triplet requires applying a microwave (MW) magnetic field at the ground-state splitting of approximately 2.87\,GHz. 
However, to fully exploit its properties, the NV must experience a strong, homogeneous MW field--a  challenge for conventional delivery schemes such as wire loops or coplanar waveguides, which typically yield fields that are either too localized or exhibit significant amplitude variation across millimeter-scales. Indeed, a further challenge lies in the fact that the ND is expected to be up to 1\,mm away from the plane of the antenna. Significant effort is being devoted to chip-compatible techniques for diamonds, for example, for sensing applications \cite{Jens_CMOS_2023}.\\

Several resonant structures have been developed. Among the earliest and most influential is the planar double split-ring resonator (DSRR) introduced by Bayat et al. in 2014 \cite{bayat_microwave_2014}. Fabricated on a PCB substrate, this device consists of two concentric copper rings, each terminated by a capacitive gap and coupled to a microstrip feed. By carefully optimizing ring radii, gap sizes, and coupling geometry, Bayat et al. demonstrated uniform MW excitation over nearly 1\,mm$^2$ of diamond, achieving Rabi frequencies exceeding 15\,MHz with only 0.5\,W of input power, corresponding to an average field of $\sim$\,5.6\,G with less than 5\,\% spatial inhomogeneity. Crucially, the DSRR design preserved optical access for fluorescence detection and was fully compatible with standard PCB fabrication techniques, making it readily adoptable for many configurations.\\

Building on this concept, Ben-Shalom et al. \cite{ben-shalom_modified_2023} introduced a vertically stacked modified split-ring resonator (MSRR) to further improve three-dimensional homogeneity, particularly along the out-of-plane (z) axis. The MSRR comprises two identical split rings separated by a few millimeters and driven either from one or two ports in phase, yielding a pronounced, uniform MW magnetic field normal to the ring planes. Experimentally, this configuration achieved Rabi frequencies up to 18\,MHz and an efficiency of 2\,${G}\over{\sqrt{W}}$  over a cylindrical volume of 0.1 mm$^3$, with field variations below 0.7\,\%. The narrow intrinsic bandwidth of $\sim$\,20\,MHz remains sufficient for many sensing protocols, while the stacked geometry retains optical access through the inter-ring gap. Moreover, the MSRR shows low sensitivity to external perturbations and can be finely tuned to the NV transition via simple metallic stubs, ensuring robust integration into complex experimental assemblies.\\

Together, these resonator designs illustrate a clear evolution in MW delivery for NV applications: from planar, in-plane uniformity (DSRR) to truly three-dimensional homogeneity (MSRR), each addressing the critical need for strong, spatially uniform drive fields over large volumes while maintaining compatibility with optical readout.\\

\begin{figure}[H]
\centering
\includegraphics[width=0.8\textwidth,trim=0mm 0mm 0mm 0mm,clip]{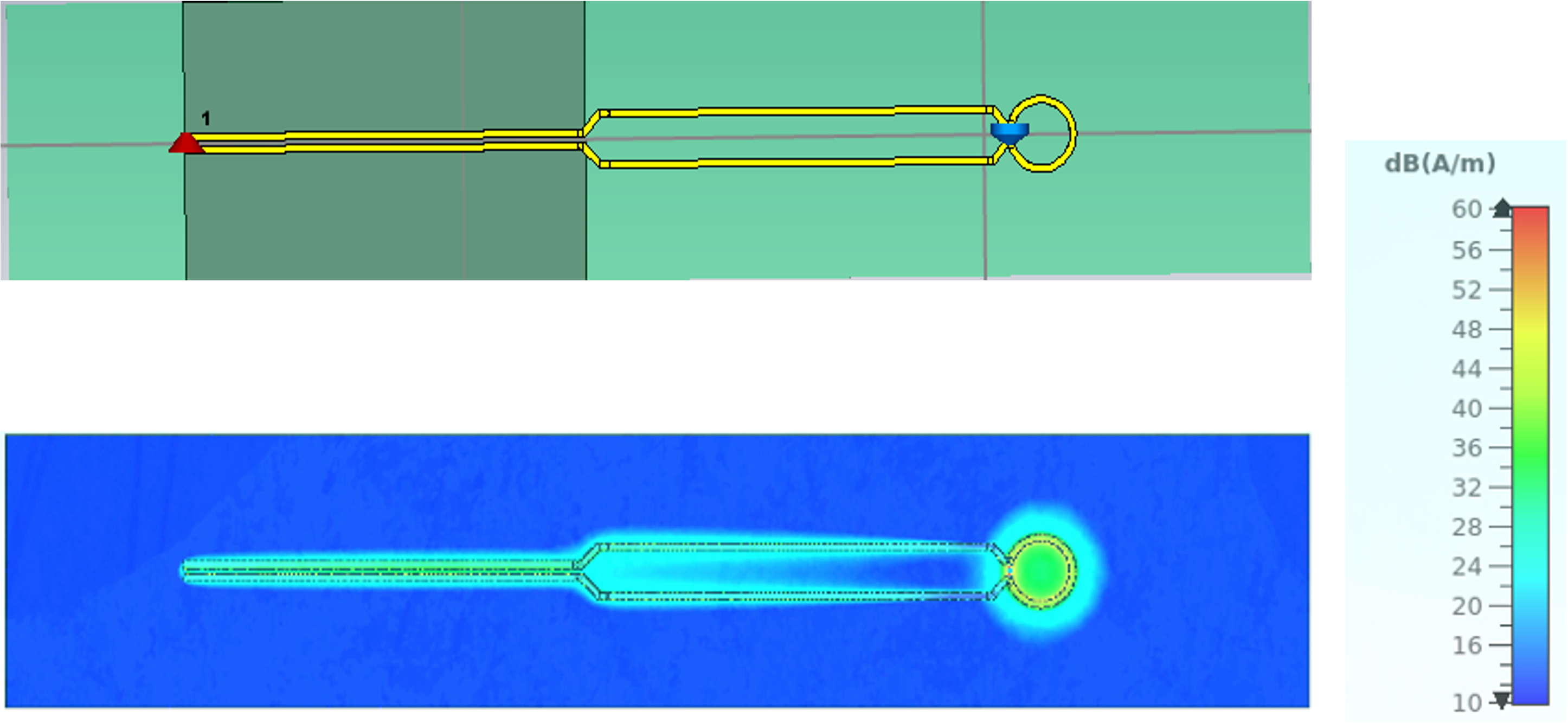}\vspace{-0.2cm}
\caption[Resonator]{(Top) An illustration of the MW resonator. (Bottom) a simulated 2D profile of the magnetic field.}
\label{fig:2Dfield.png}
\end{figure}

The goal for our Generation-1 antenna is to develop a tunable microwave resonator that produces a magnetic field in the 2.87 GHz band precisely matching the electron spin resonance of NV centers in diamond and to deliver field amplitudes between 1\,G and 5\,G over a bandwidth of roughly 300\,MHz to accommodate Zeeman shifts under varying external fields.\\

\begin{figure}[H]
\centering
\includegraphics[width=0.8\textwidth,trim=0mm 0mm 0mm 0mm,clip]{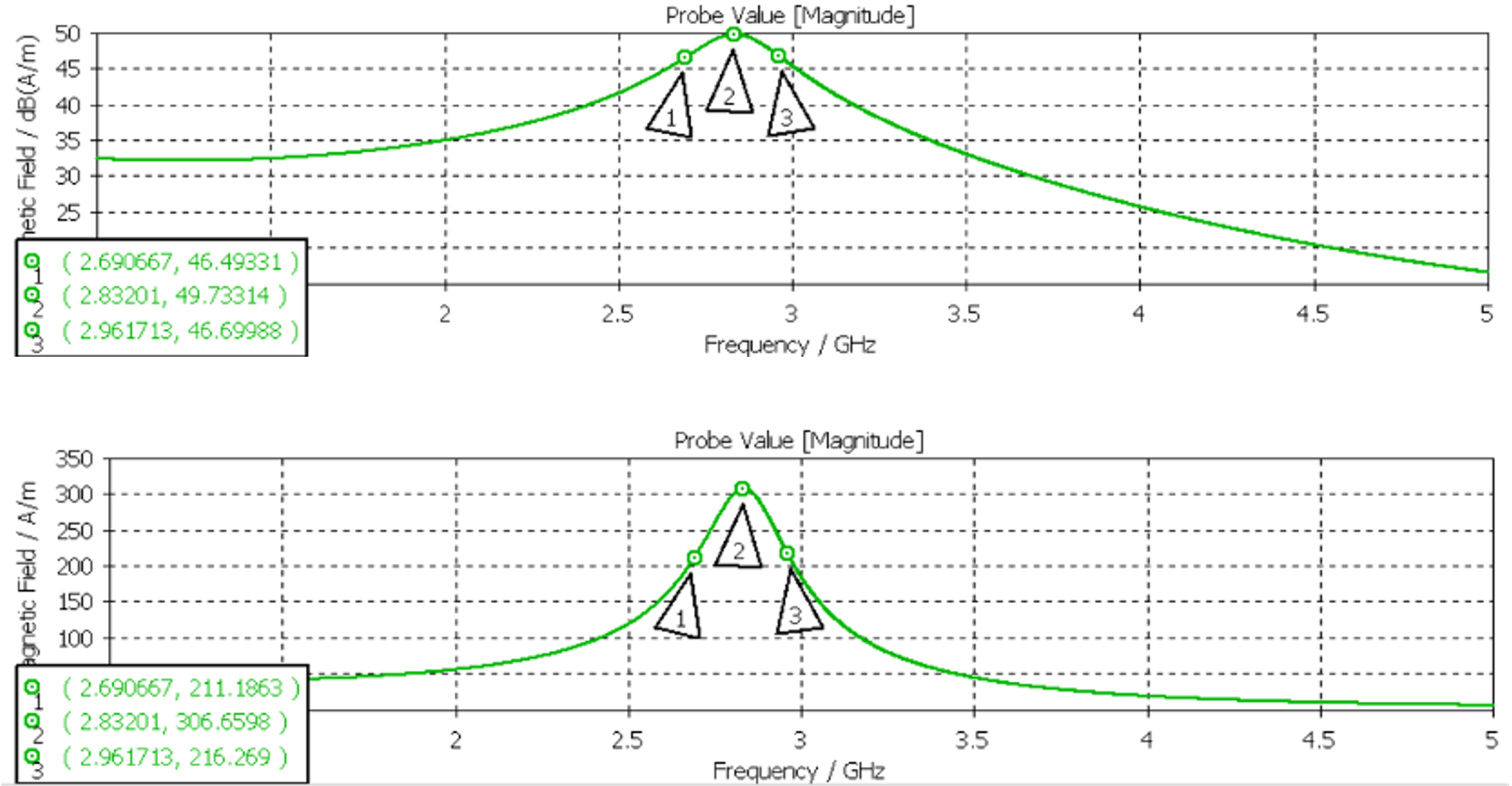}\vspace{-0.2cm}
\caption[SimAntenna]{Simulated magnetic field profile in the center of the resonator's ring.}
\label{fig:Field_Profile.png}
\end{figure}

\begin{figure}[H]
\centering
\includegraphics[width=0.8\textwidth,trim=0mm 0mm 0mm 0mm,clip]{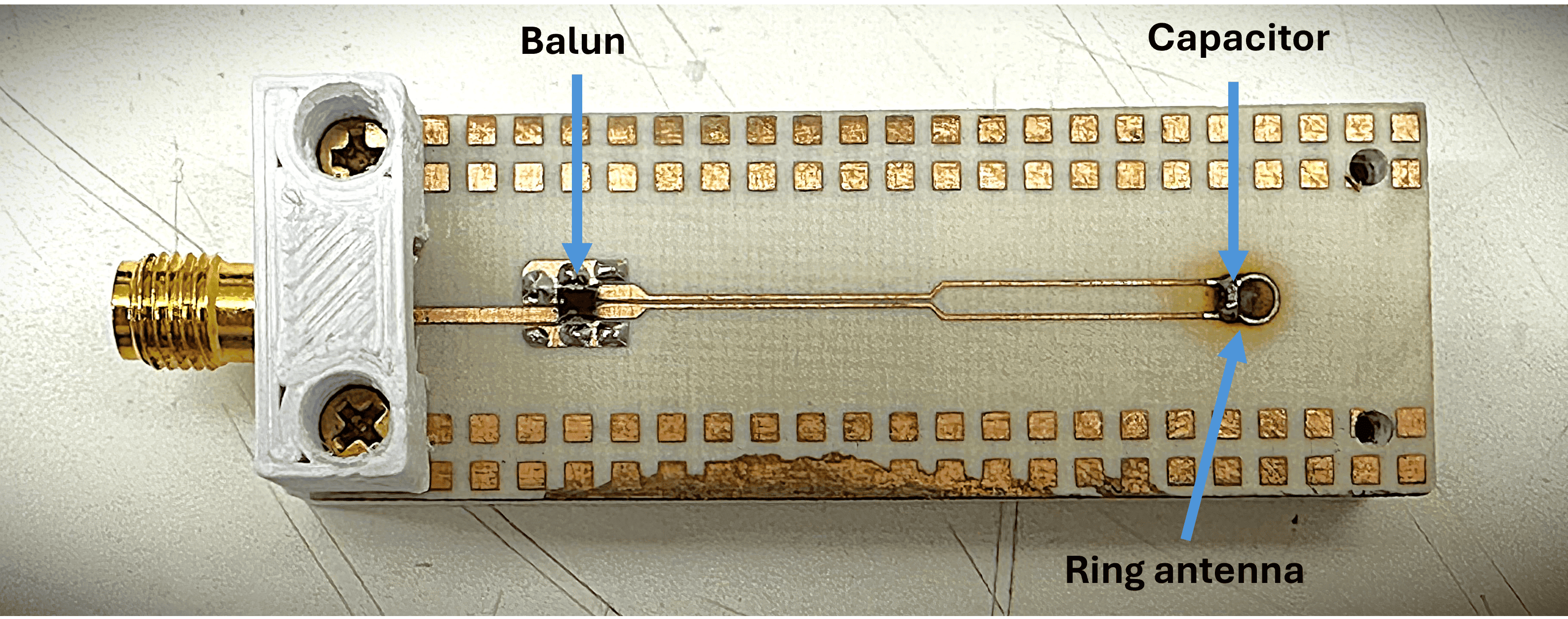}\vspace{-0.2cm}
\caption[Antenna]{The MW resonator we used in our experiments.}
\label{fig:AntennaReal.png}
\end{figure}

To achieve this, we implemented a parallel resonator circuit whose resonant enhancement amplifies current and thus the generated magnetic field, when capacitive and inductive reactance's cancel. By carefully selecting the series resistance, we set the bandwidth to our 300\,MHz target. Because the inductor must see a balanced RF drive (currents entering and exiting inductor windings), we introduced a balun to convert the microwave source’s 50\,$\Omega$ unbalanced output into the needed differential impedance. Impedance transformations were carried out using quarter-wavelength transmission lines: stepping 100\,$\Omega$ to 650\,$\Omega$ via coupled microstrip sections. All electromagnetic and circuit elements, including the split-ring inductor and the transmission line networks were modeled and optimized in CST Studio Suite to ensure accurate resonant frequency, Q-factor, and impedance matching (see Fig.\,\ref{fig:2Dfield.png}). 
In its final form, the resonator produces a peak magnetic field of approximately 3.76\,G at the NV ESR frequency, with a measured 3\,dB bandwidth of about 270\,MHz and spatial uniformity of $\pm$ 6.6\,\% over 1\,mm$^2$ when driven with 0.5\,W of input power closely meeting our design targets for amplitude tunability and spectral coverage (see Fig.\,\ref{fig:Field_Profile.png}). In Fig.\,\ref{fig:AntennaReal.png} we show an image of the device.\\

In summary, we have realized a compact, broadband microwave resonator that delivers uniform, tunable magnetic fields precisely at the NV center’s ESR frequency. By leveraging a high-Q RLC topology with tailored impedance-transform networks, the device meets the target field amplitudes and spectral bandwidth, while maintaining excellent impedance matching and optical access. Its straightforward design and reproducible performance make it advantageous not only for the ND SGI, but also a versatile platform for ensemble NV magnetometry and broader quantum-sensing applications.

\section{APPENDIX B.\,\,\, Hyperfine splitting}

Here we explain the theory behind Fig.\,\ref{fig:RamseyVsTime.png}. Specifically we explain why we expect to observe three close transitions spaced by about 2\,MHz. In our explanation we follow Fig.\,\ref{fig:Hyperfine_Splittings.png}.\\

The Hamiltonian of the NV system may be divided into three, where the first, $H_S$, is related to the electron spin, the second, $H_I$, related to the nuclear spin, and the third, $H_{SI}$, related to the interaction between the spins. Specifically, there are
\begin{equation}
H_S = D_{gs}S_z^2 + g_s \mu_B (\mathbf{B} \cdot \mathbf{S}) + E(S_y^2-S_x^2)\,,
\end{equation}
\vspace{-0.5em}
\begin{equation}
H_I = PI_z^2 - g_I \mu_N (\mathbf{B} \cdot \mathbf{I})\,,
\end{equation}
\vspace{-0.5em}
\begin{equation}
H_{SI} = A(\mathbf{S} \cdot \mathbf{I})\,,
\end{equation}
where the latter may also be written as $H_{SI}=A_{\parallel} I_z\cdot S_z + A_{\perp}(I_x\cdot S_x+I_y\cdot S_y)$, reducing to just the first term in the case of an aligned system. The second Hamiltonian reduces to $H_I=PI_z^2$ when the magnetic field is small, as $\mu_N$ is smaller than $\mu_B$ by a factor of almost 2000. The values for $P$ and $A_{\parallel}$ are -4.95\,MHz and 2.16\,MHz, respectively.\\

The expected splitting between the hyperfine levels is noted in the left plot of Fig.\,\ref{fig:Hyperfine_Splittings.png}. Taking into account both the splitting in the ground state and the splitting in the excited state, the three transitions to one of the electronic $m_s=\pm 1$ states are spaced by about 2\,MHz to each side of the $m_I=0$ transition.

\section*{Acknowledgments}
We thank the BGU Atom-Chip Group support team, especially Menachem Givon, Zina Binstock, Dmitrii Kapusta, Itamar Haskel and Yaniv Bar-Haim for their support in building and maintaining the experiment. Funding: This work was funded by the Gordon and Betty Moore (doi.org/10.37807/GBMF11936), and Simons (MP-TMPS-00005477) Foundations.

\end{document}